\documentclass[11pt]{article}
\usepackage{fullpage}

\newtheorem{theorem}{Theorem}[section]

\def\square{\hfill\nobreak\rule{1ex}{1.4ex}}

\def\AC0{{\rm AC}\,{^0}}
\def\CC0{{\rm CC}\,{^0}}
\def\ACC0{{\rm ACC}\,{^0}}
\def\NC1{{\rm NC}\,{^1}}

\def\Sig_#1{{\Sigma^p_#1}}
\def\Deltap_#1{{\Delta^p_#1}}
\def\Pip_#1{{\Pi^p_#1}}

\long\def\COMMENT#1{\relax}   

\begin{document}
\sloppy

\title{Quantum Search Algorithms}
%
\medskip
\author{Andris Ambainis\footnote{Institute for Quantum Computing 
and Department of
Combinatorics and Optimization, University of Waterloo, 
e-mail: {\tt ambainis@math.uwaterloo.ca} This work done at 
School of Mathematics, Institute for Advanced Study, 
Princeton, NJ 08540, USA. 
Supported by NSF Grant DMS-0111298. Any opinions, findings and conclusions 
expressed in this material are those of the author and do not necessarily reflect views of the National Science Foundation.}}

\date{}

\maketitle

\begin{abstract}
We review some of quantum algorithms for search problems:
Grover's search algorithm, its generalization to amplitude amplification,
the applications of amplitude amplification to various problems and
the recent quantum algorithms based on quantum walks. 
\end{abstract}

\section{Introduction}

Quantum computation explores the possibilities of applying quantum mechanics
to computer science. 
If built, quantum computers would provide speedups over conventional
computers for a variety of problems.
The two most famous results in this area are 
Shor's quantum algorithms for factoring and finding 
discrete logarithms \cite{Shor94} and Grover's search algorithm \cite{Grover96}.

Shor's and Grover's algorithms have been followed by a lot of other results.
Each of these two algorithms has been generalized and applied to several other
problems. New algorithms and new algorithmic paradigms (such as adiabatic 
computing\cite{Farhi} which is the quantum counterpart of simulated annealing)
have been discovered.

In this column, we survey some of the results on quantum algorithms,
focusing on the branch of quantum algorithms inspired by
Grover's search algorithm \cite{Grover96}. 

Instead of the conventional introduction/review on quantum computing which starts 
with the backgrounds from physics, we follow a different path. 
We first describe Grover's search result and 
its generalization, {\em amplitude amplification} (section \ref{sec:grover}).
Then, we explore what can be obtained by using these results
as ``quantum black boxes" in a combination with methods from 
conventional (non-quantum) algorithms and complexity (section \ref{sec:apps}).
We give three examples of quantum algorithms of this type, 
one very simple and two more advanced ones.   
After that, in section \ref{sec:qw}, we show 
some examples were simple application of Grover's search fails 
but more advanced quantum algorithms (based on quantum walks) succeed.

\section{Grover's search and amplitude amplification}
\label{sec:grover}

Grover's search algorithm is one of main quantum algorithms. 
The problem that it solves is very simple to state:

{\bf Search.} We have an input $x_1, \ldots, x_N\in\{0, 1\}$ 
specified by a black box that answers queries. 
In a query, we input $i$ to the black box and it outputs $x_i$. 
Our task\footnote{There are several variations of this problem.
We could either require the algorithm to output ``none'' if there is 
no $i:x_i=1$ or allow any output in this situation.
Or we could consider a decision version, 
where the algorithm only has to determine if there exists
$i:x_i=1$ instead of finding $i$. 
The complexity remains almost the same for all variations.} is to output
an $i:x_i=1$.

Then, $N$ queries are needed for deterministic algorithms
and $\Omega(N)$ queries are needed for probabilistic algorithms.
(This follows by considering the case when there is 
exactly one $i$ such that $x_i=1$ and $N-1$ variables $i$:$x_i=0$.)

Grover \cite{Grover96} studied the quantum version of this problem 
(in which the black box is quantum, the input to the black box 
is a quantum state consisting of various $i$ and the output
is the input state modified depending on $x_i$). His result is

\begin{theorem}
\cite{Grover96}
{\bf Search} can be solved with $O(\sqrt{N})$ quantum queries.
\end{theorem}

\subsection{Is it ``database search''?}
\label{sec:db}

Grover's algorithm is often called "database search". 
In this interpretation, the variables $x_1$, $\ldots$, $x_N$ 
correspond to $N$ entries of the database. A variable is 1 if
the corresponding entry of database matches our search criteria.
Then, the result is that we can search an unordered database 
of $N$ entries in time $O(\sqrt{N})$.

This interpretation has caused some heated debates.
Essentially, the issue is that, to access $N$ elements,
we would need quantum hardware of size $\Omega(N)$ 
\cite[section 6.5]{NC}.
Since quantum hardware is likely to be expensive,
this may be a big obstacle.  

Nevertheless, Grover's algorithm can be very useful in
problems of a different nature. 
Say we have an instance of an NP-complete problem, for example,
satisfiability.
That is, we have a boolean formula $F(y_1, \ldots, y_n)$ 
and we want to know if one of $2^n$ assignments 
$y=(y_1, \ldots, y_n)$ of values to the variables makes $F$ true. 
The naive exhaustive search requires testing $2^n$ assignments. 

With a quantum computer, we could instead reduce the satisfiability
to search on $N=2^n$ variables $x_1, \ldots, x_N$ with $x_i=1$ if
$F$ is true for the $i^{\rm th}$ candidate assignment 
$y=(y_1, \ldots, y_n)$. The black box that answers queries
is just a circuit that takes an assignment $(y_1, \ldots, y_n)$ and
checks if $F$ is true on this assignment. Then, Grover's algorithm
allows to solve satisfiability in time $O(\sqrt{N})=O(1.41...^n)$ 
instead of $N=2^n$. Similar approach applies to any exhaustive 
search problem.

A knowledgeable reader might point out that 3-SAT can be solved 
even faster classically, in time $O(1.329...^n)$ by a non-naive 
algorithm \cite{Schoning,3SAT,Rolf}. 
We address this issue in section \ref{sec:3-sat}.

\subsection{Facts about Grover's algorithm}

Since Grover's result, the search problem has been analyzed in great detail.
Here are some of results that we know:
\begin{enumerate}
\item
In general, Grover's algorithm is bounded-error. Given a black-box 
$x_1, \ldots, x_N\in\{0, 1\}$ where some $x_i$ are equal to 1, the
algorithm might not find any of them with a small probability.
However, if we know that the number of $i:x_i=1$ is exactly $k$, 
then the algorithm can be tuned so that it finds one of them 
with certainty (probability 1) in $O(\sqrt{N/k})$ steps \cite{BHMT}. 
\item
Moreover, if we know that the number of $i:x_i=1$ is exactly $k$,
the algorithm is {\em exactly} optimal \cite{Zalka}.
The number of queries cannot be improved even by 1.
For finding $i:x_i=1$ with certainty, 
the minimum number of queries is known to be exactly
\begin{equation}
\label{eq:steps} 
\left\lceil \frac{\pi}{4\arcsin \frac{1}{\sqrt{N/k}}} - \frac{1}{2}
\right\rceil < \frac{\pi}{4} \sqrt{\frac{N}{k}} 
\end{equation}
If the number of queries $t$ is less than that, the best probability
with which any quantum algorithm can find an $i:x_i=1$ is exactly
the one achieved by running Grover's algorithm with $t$ queries.
\item
If $k$ is unknown, $O(\sqrt{N})$ queries are still sufficient.
If $k$ is unknown but it is known that $k\geq k_0$, $O(\sqrt{N/k_0})$ queries
suffice \cite{TightSearch}.
\item
In this case, the algorithm is inherently bounded-error.
There is no quantum algorithm with less than $N$ queries that solves
Grover's problem with certainty for arbitrary $x_1, \ldots, x_N$ \cite{Beals}.  
\end{enumerate}

If we have an instance $x_1, \ldots, x_N$ with $k$ elements equal to 1 and would
like to find all $k$ of them, $\Theta(\sqrt{Nk})$ queries are sufficient and necessary.

\subsection{Amplitude amplification}
\label{sec:ampl}

Let $A$ be a (classical or quantum) algorithm with one sided error.
If the correct answer is ``no'', $A$ always outputs ``no''.
If the correct answer is ``yes'', $A$ outputs ``yes'' with
at least some (small) probability $\epsilon>0$. 

An example is an algorithm for SAT (or any other problem in $NP$) 
which outputs ``the formula is satisfiable''
only if it finds a satisfying assignment.
Or, a different example is an algorithm for Grover's search
problem which outputs ``there exists $i:x_i=1$'' only if it has
found such $i$.

How many times do we need to repeat the algorithm to increase its
success probability from a small $\epsilon$ to a constant (for example, 2/3)?
algorithm to increase the success probability? Classically, 
$\Theta(1/\epsilon)$ repetitions are needed. In quantum case,
a generalization of Grover's algorithm gives

\begin{theorem}
\cite{BHMT}
Let $A$ be a quantum algorithm with one-sided error and success 
probability at least $\epsilon>0$.
Then, there is a quantum algorithm $B$ that solves the same
problem with success probability 2/3 by invoking 
$A$ $O(\frac{1}{\sqrt{\epsilon}})$ 
times.
\end{theorem}

This result is called {\em amplitude amplification}.
For more details, see \cite{BHMT}.
Similar result is also known for algorithms with two-sided 
error but it has not found as many applications as 
amplitude amplification for algorithms with one-sided error.

\section{Three applications}
\label{sec:apps}

In this section, we show 3 examples how Grover's algorithm and 
amplitude amplification can be used to solve other problems. 
The examples are selected to be solvable by just two ideas
from quantum computation (and some algorithmic ingenuity).
The first of two ideas is Grover's search and amplitude amplification,
described in the previous section.
The second idea is that any classical (either 
deterministic or probabilistic) computation
can be simulated on a quantum computer \cite[section 1.4]{NC}.
More precisely, 
\begin{itemize}
\item
In the circuit model, a classical circuit with $N$ gates
can be simulated by a quantum circuit with $O(N)$ gates.
\item
If the query model (when only the number of queries is counted),
a classical computation with $N$ queries can be simulated 
by a quantum computation with $N$ queries.
\end{itemize}
This greatly simplifies descriptions of quantum algorithms.
Instead of describing a quantum algorithm, we can describe
a classical algorithm that succeeds with some small 
probability $\epsilon$. Then, we can
transform the classical algorithm to a quantum algorithm
and apply the amplitude amplification to the quantum
algorithm. The result is a quantum algorithm with the running time or
the number of queries that is $O(\frac{1}{\sqrt{\epsilon}})$
times the one for the classical algorithm with which we started.

A similar reasoning can be applied, if instead of a 
purely classical algorithm, we started with a classical algorithm
that involves quantum subroutines. Such algorithms can also
be transformed into quantum algorithms with the same complexity.

\subsection{3-satisfiability}
\label{sec:3-sat}

As we described in section \ref{sec:db},
Grover's algorithm can solve 3-satisfiability in 
$O(1.41...^n poly(n))$ steps.
However, the best known classical algorithm for 3-satisfiability is faster than that, running 
in time $O(1.329...^n)$ \cite{Rolf}. Does this mean that Grover's algorithm is not useful
for satisfiability?

Not quite. The best classical algorithm can be combined 
with Grover's search. The result is a quantum algorithm that runs in
time $O(1.153...^n poly(n))$, providing a square-root-speedup over \cite{Rolf}.

We first describe the classical algorithm (due to Sch\"oning \cite{Schoning},
improved by \cite{3SAT,Rolf}).
Its structure is as follows:
\begin{enumerate}
\item
Pick a random initial assignment $x_1, \ldots, x_n$.
\item
$3n$ times repeat:
\begin{enumerate}
\item
If all clauses satisfied, stop.
\item
Otherwise, find an unsatisfied clause. Make it satisfied by choosing a variable in this clause
uniformly at random and changing its value.
\end{enumerate}
\end{enumerate}
The result of \cite{Rolf} is that, for an appropriate initial probability
distribution in the first step,
the algorithm finds a satisfying assignment 
(if there is one) with a probability at least $c^n$ (where 
$c=\frac{1}{1.329...}$).
Repeating the algorithm $1.329...^n$ times 
gives an algorithm that finds a satisfying
assignment in time $O(1.329...^n poly(n))$ with a constant success probability.

To obtain a quantum algorithm, we just use quantum amplitude amplification instead 
of classical repetition. As described in section \ref{sec:ampl}, amplitude amplification
allows to increase the success probability to a constant, repeating the algorithm
$O(\frac{1}{\sqrt{\epsilon}})$ times. In this case, this means 
$O(\sqrt{1.329...^n})=O(1.153...^n)$ repetitions.

The result is very simple but it illustrates an important point.
For some problems, Grover's algorithm can provide a quadratic speedup not
just over the naive classical algorithm (testing all assignments) but over
better classical algorithms as well.

\subsection{Element distinctness}
\label{sec:ed}

{\bf Element Distinctness.}
We are given $f:\{1, 2, \ldots, N\}\rightarrow\{1, 2, \ldots, N\}$
specified by a black box that, given $i$, answers the value of $f(i)$.
The task is to determine if there are
two inputs $i, j$, $i\neq j$ for which $f(i)=f(j)$.

The measure of complexity is the number of queries to the black box. 
Classically, this problem requires $\Omega(N)$ queries. In quantum case,
there are two algorithms. The first, due to \cite{Buhrman01} uses Grover search in 
a clever two-level construction and solves the problem $O(N^{3/4})$ queries. 
The second, due to \cite{Ambainis04}, uses a technique combining search with
quantum walks and solves the problem with $O(N^{2/3})$ queries. 
This is optimal, because of an $\Omega(N^{2/3})$ lower bound by Shi \cite{Shi}. 

In this section, we show the $O(N^{3/4})$ algorithm by Buhrman et.al. \cite{Buhrman01}.
While the result is weaker than the later algorithm of \cite{Ambainis04}, the idea is
very elegant.
Consider the following algorithm:
\begin{enumerate}
\item
Choose $\sqrt{N}$ random numbers $i_1, \ldots, i_{\sqrt{N}}\in\{1, 2, \ldots, N\}$.
Evaluate $f(i_1)$, $\ldots$, $f(i_{\sqrt{N}})$. 
If two of them are equal, stop, output the two equal elements.
\item
Use Grover's search to search (among remaining $N-\sqrt{N}$ indices
$k\in\{1, 2, \ldots, N\}$) for an index $k$ such that $f(k)=f(i_j)$ for some $j$. 
\end{enumerate}
This algorithm requires $\sqrt{N}$ queries for the first step and 
$O(\sqrt{N})$ queries for the second step. (Notice that we do not need to
query $f(i_j)$ since their values are known from the first step.)
The total number is $O(\sqrt{N})$.

If there is a pair $i, j$ such that $f(i)=f(j)$, then, with probability 
$\frac{\sqrt{N}}{N}=\frac{1}{\sqrt{N}}$, $i$ is among 
$i_1, \ldots, i_{\sqrt{N}}$. In this case, the second step will find $j$ 
(or some other element $k$ such that $f(k)$ is equal to one 
of $f(i_1)$, $\ldots$, $f(i_{\sqrt{N}})$)
a constant probability. Thus, the algorithm succeeds with probability
at least $\frac{const}{\sqrt{N}}$. 

We can now apply the amplitude amplification, described in
section \ref{sec:ampl}. It increases the success probability to a constant
with $O(\frac{1}{\sqrt{\epsilon}})$ repetitions of the whole algorithm.
Since $\epsilon=\frac{const}{\sqrt{N}}$, $O(N^{1/4})$ repetitions
suffice. The total number of queries needed is 
$O(N^{1/4} N^{1/2})=O(N^{3/4})$.

\subsection{Finding global and local minima}

In this section, we describe quantum algorithms for two minimum-finding
problems. 

{\bf Global Minimum.}
We have an integer-valued function $f(i)$ of one variable $i\in\{1, 2, \ldots, N\}$,
specified by black box that answers queries.
The input of a query is $i\in\{1, 2, \ldots, N\}$, the output is
$f(i)$. The task is to find $i$ such that $f(i)\leq f(j)$ for any $j\neq i$.

{\bf Local Minimum.}
We have an integer-valued function $f(x_1, \ldots, x_n)$ of Boolean variables 
$x_1, \ldots, x_n \in\{0, 1\}$, specified by black box that answers queries.
The input of a query is $x_1, \ldots, x_n \in\{0, 1\}$, the output is
$f(x_1, \ldots, x_n)$. 
The task is to find a local minimum: an assignment
$x_1, \ldots, x_n$ such that changing any one variable does not decrease the value
of the function:
\[ f(x_1, \ldots, x_{i-1}, 1-x_i, x_{i+1}, \ldots, x_n)\geq 
f(x_1, \ldots, x_n) .\]

In both cases, the measure of complexity is the number
of queries (i.e. the number of times that we need to evaluate $f$).  
We start by describing an algorithm for the first problem, which
will be used as a subroutine in the second algorithm.

\begin{theorem}
\cite{Durr}
{\bf Global Minimum} can be solved with $O(\sqrt{N})$ quantum queries.
\end{theorem}

Classically, $\Omega(N)$ queries are required.

The outline of the algorithm is as follows (some technical details are omitted,
to simplify the presentation):
\begin{enumerate}
\item
Choose $x$ uniformly at random from $\{1, \ldots, N\}$; 
\item
Repeat:
\begin{enumerate}
\item
Use Grover's search to search for $y$ with $f(y)<f(x)$;
\item
If search succeeds, set $x=y$. Otherwise, stop and output $x$ as the minimum.
\end{enumerate}
\end{enumerate}

We sketch why $O(\sqrt{N})$ queries are sufficient for this algorithm,
on intuitive but ``hand-waving" level.
(For a more detailed and rigorous argument, see \cite{Durr}.) 
For simplicity, assume that, for all $x$, the values of $f(x)$ are distinct. 
Let $x_0$ be the value of $x$ at the beginning of the algorithm and $x_i$
be the value of $x$ after the $i^{\rm th}$ Grover's search. 
Since $x_0$ is a random element of $\{1, \ldots, N\}$,
$f(x_0)$ will, on average, be the $(N/2)^{\rm th}$ smallest
element of $\{f(1), \ldots, f(N)\}$.
After the first iteration, $x_1$ is some element with $f(x_1)<f(x_0)$.
By inspecting Grover's algorithm, we can find out that the probabilities
of algorithm outputting $x_1$ are equal for all $x_1$ with $f(x_1)<f(x_0)$.
Thus, $x_1$ is uniformly random among numbers with $f(x_1)<f(x_0)$.
Since $f(x_0)$ was, on average, be the $(N/2)^{\rm th}$ smallest
element of $\{f(1), \ldots, f(N)\}$, this means that $f(x_1)$ is, 
on average, the $(N/4)^{\rm th}$ smallest element. 
By a similar argument, $f(x_i)$ is, on average the $(N/2^i)^{\rm th}$ 
smallest element in $\{f(1), \ldots, f(N)\}$.

We now remember that Grover's search uses $O(\sqrt{N/k})$ queries
where $k$ is the number of solutions. Consider repetitions of the minimum
finding algorithm in the order from the last to the first.
By the argument above, we would expect that, in the last iteration before finding the
minimum, $k\approx 1$, then, in the iteration before that, $k\approx 2$,
then $k\approx 4$ and so on. Then, the total number of queries in all
the repetitions of Grover's search is of order
\begin{equation}
\label{eq:min}
 \sqrt{N} + \sqrt{N/2}+\sqrt{N/4}+ \ldots = \sqrt{N}
\left(1+ \frac{1}{\sqrt{2}}+\frac{1}{2}+\ldots \right) .
\end{equation}
The term in brackets is a decreasing geometric progression and,
therefore, sums up to a constant. 
This means that the sum of equation (\ref{eq:min}) is of order $O(\sqrt{N})$.

We now turn to algorithms for {\bf Local Minimum}.
Classically, $\Theta(2^{n/2}poly(n))$ 
queries are necessary and sufficient \cite{Aldous,Aaronson03}. In quantum case,

\begin{theorem}
\cite{Aaronson03}
{\bf Local minimum} can be solved with $O(2^{n/3}n^{1/6})$ quantum queries.
\end{theorem}

The algorithm (again, in a simplified form) is as follows:
\begin{enumerate}
\item
Choose $m$ assignments $x=(x_1, \ldots, x_n)$ uniformly at random. 
Use Grover's search to find one with the smallest value of $f(x_1, \ldots, x_n)$.
\item
$2^{n+1}/m$ times repeat: 
\begin{enumerate}
\item
Use Grover's search
to search for $(y_1, \ldots, y_n)$ with
$f(y_1, \ldots, y_n)<f(x_1, \ldots, x_n)$, among $n$ assignments 
$(y_1, \ldots, y_n)$ that differ from 
$(x_1, \ldots, x_n)$ in exactly one variable.
\item
If such $(y_1, \ldots, y_n)$ is found, set 
$(x_1, \ldots, x_n)=(y_1, \ldots, y_n)$.
Otherwise, stop and claim that $(x_1, \ldots, x_n)$ is a local minimum.
\end{enumerate}
\end{enumerate}

The first step requires $O(\sqrt{m})$ queries. 
The second step requires $O(\sqrt{n})$ 
queries each time it is repeated. 
The total number is
$O(\sqrt{m}+\frac{2^n}{m}\sqrt{n})$.
The minimum of this expression is $O(2^{n/3}n^{1/6})$ 
which is achieved by setting $m=2^{2n/3}n^{1/3}$.

The correctness of the algorithm follows from the fact that, if we pick $m$ elements out of
$2^{n}$, then, with high probability, one of those $m$ elements will be among $2^{n+1}/m$ smallest
among all $2^{n}$ elements (see \cite{Aaronson03} for a proof). If this is the case, the minimum
of $m$ elements is also among $2^{n+1}/m$ smallest elements
among all $2^{n}$ elements. Then, the second step of the algorithm will lead
to a local minimum in at most $2^{n+1}/m$ steps because each step replaces
$(x_1, \ldots, x_n)$ by an assignment with a smaller value of $f$ and there are
at most $2^{n+1}/m$ assignments for which $f$ has smaller value than for
the starting point.


\section{Local search and quantum walks}
\label{sec:qw}

\subsection{Two problems}

Next, we 
show two situations when a simple application
of Grover's algorithm does not give a good quantum algorithm.

\noindent
{\bf Search on grid.}
Consider $N$ memory cells, arranged into $\sqrt{N}\times\sqrt{N}$
grid. Each cell stores an element $x_i\in\{0, 1\}$. Our task is to
find an element $i:x_i=1$. At each moment of time, we are in some
memory location. In one time step, we can either query the current 
location or move to an adjacent cell. 
(In the quantum version, we can be in a quantum state consisting of
various locations. But we still require that no part of this state
moves more than distance 1 in one time unit.)

Grover's algorithm finds an element $i:x_i=1$ with $O(\sqrt{N})$ queries.
But, between any two queries, it needs $\Theta(\sqrt{N})$
moves, since a query to one element can be followed by a
query to any other element. The total number of steps is of
order $\sqrt{N}\times\sqrt{N}=N$. A similar number of steps
can be achieved by a classical algorithm that just traverses  
the grid row by row and queries every cell. 
That takes $O(N)$ moves and $O(N)$ queries, for a total of $O(N)$ 
steps as well. The quantum advantage seems to disappear \cite{Benioff}.

If the $N$ items are arranged in 3 dimensions, in a cube with side 
of length $O(\sqrt[3]{N})$, then the straightforward quantum search
takes $O(\sqrt{N}\sqrt[3]{N})=O(N^{5/6})$ which is better than classical
$O(N)$ but still worse than $O(\sqrt{N})$ in the usual Grover's search 
when only queries are counted.

\noindent
{\bf Searching the element distinctness graph.}
Element distinctness reduces to search a certain graph.
Let $1\leq M<N$. Define a bipartite graph, with the vertices
being all subsets of $\{1, 2, \ldots, N\}$ of size $m$ and size $m+1$.
A vertex $v_S$ corresponding to a subset $S$ is connected 
to a vertex $v_T$ if $|S|=M$, 
$|T|=M+1$ and $T=S\cup\{i\}$ for some $i\in \{1, 2, \ldots, N\}$.
A vertex $v_S$ is marked if the set $S$ contains $i, j$ such that $i\neq j$ 
and $x_i=x_j$. The task is to find a marked vertex. In one step, we are allowed
to examine the current vertex or to move to an adjacent vertex.

If we can search this graph in $f(N)$ steps, we can solve element
distinctness with at most $f(N)+M$ queries, in a following way. 
If we are at a vertex $v_S$, we will know the values of all $x_i$, $i\in S$.
Then, testing if a vertex is marked can be done with no queries and moving
to an adjacent vertex $v_T$ can be done with 1 query
by querying the only element $i\in T-S$. 
To achieve that, we use the first $M$ queries to query all $x_i$ for $i\in S$ where $S$ 
is the set corresponding to the starting vertex $v_S$. 
The total number of queries is $M$ to start the algorithm and 
at most one query per search algorithm step afterwards.

Again, let us try to use Grover's algorithm to search this graph.
If there is exactly one pair of equal elements $x_i=x_j$, then the probability
of a random vertex $v_S$, $|S|=M$ being marked is
\[ Pr[i\in S] Pr [j\in S|i\in S]= \frac{M}{N} \frac{M-1}{N-1}=(1+o(1)) \frac{M^2}{N^2} \]
Amplitude amplification
implies that we can find such set $S$ by testing $O(\frac{N}{M})$ vertices
which is a square root of what one would need classically.
However, testing each vertex $v_S$ involves querying $M$ elements.
The total number of queries is of order $\frac{N}{M} M =N$
which is the same as if we just queried all $N$ elements $x_i$ 
to begin with.

Both examples illustrate the same general situation. 
Sometimes, we have a search space, 
where after testing an item, it is faster to test
a neighboring item than an arbitrary item. This could come either from
physical constraints on the search space (the first example) or
an algorithmic structure (the second example). A straightforward
application of Grover's algorithm does not do well on such spaces,
because it does not respect the structure of the space.

\subsection{Solution to two problems}

There is a recent approach, based on quantum walks (quantum counterparts
of random walks) that overcomes this problem. 
(For more information on quantum walks, see the surveys
\cite{Kempe,Ambainis03}.)

\begin{theorem}
\label{th:akr}
\cite{AKR}
{\bf Spatial search} can be solved with 
\begin{enumerate}
\item
$O(\sqrt{N}\log N)$ steps in 2 dimensions, if there is a unique $i:x_i=1$;
\item
$O(\sqrt{N}\log^2 N)$ steps in 2 dimensions in the general case (no assumptions on 
the number of $i:x_i=1$);
\item
$O(\sqrt{N})$ steps in 3 and more dimensions.
\end{enumerate} 
\end{theorem}

Quantum walks also give a better search algorithm for the element
distinctness graph.
It can be searched in $O(N/\sqrt{M})$ steps, implying an algorithm
for element distinctness with $O(M+\frac{N}{\sqrt{M}})$ steps.
Setting $M=N^{2/3}$ gives

\begin{theorem}
\label{th:ed}
\cite{Ambainis04}
{\bf Element distinctness} can be solved with $O(N^{2/3})$ queries.
\end{theorem}

For the first result (spatial search), there is a different solution
that involves amplitude amplification instead of quantum walks \cite{AA},
giving $O(\sqrt{N})$ in 3 dimensions and $O(\sqrt{N}\log^2 N)$ 
and $O(\sqrt{N}\log^3 N)$ in the two 2-dimensional cases. For element
distinctness, quantum walks are the only approach 
known to give $O(N^{2/3})$ algorithm.

Szegedy \cite{Szegedy} has generalized element distinctness and
spatial search, by showing how to convert a general
classical Markov chain into a quantum walk algorithm.
His generalization of element distinctness is

\begin{theorem}
\label{th:szegedy}
\cite{Szegedy}
Let $P$ be a symmetric Markov chain with a gap between the first
and the second eigenvalue being $\epsilon$ and
at least $\delta$ fraction of states being marked. 
Assume that we can perform the following operations:
\begin{itemize}
\item
generate a uniformly random element in $\gamma_0$ steps;
\item
given a state $x$, generate a sample from $P(x, y)$ in $\gamma_1$ steps;
\item
check if a state is marked in $\gamma_2$ steps.
\end{itemize}
Then, there is a quantum algorithm that finds a marked state
in $O(\gamma_0+\frac{1}{\sqrt{\delta\epsilon}} (\gamma_1+\gamma_2))$ steps.
\end{theorem}

For comparison, a classical random walk would find a marked state in
$O(\gamma_0+\frac{1}{\delta\epsilon} (\gamma_1+\gamma_2))$ steps.

The element distinctness algorithm is a special case of Theorem \ref{th:szegedy}
where the states of Markov chain are sets $S$, $|S|=M$ or $|S|=M+1$ and, at each step,
the Markov chain adds (if $|S|=M$) or removes (if $|S|=M+1$) a random
element from the set. 

\subsection{Applications of element distinctness}

There are several results that build on element distinctness algorithm
\cite{MSS,CE,BS}.

{\bf Triangle finding.} 
A graph $G$ with $N$ vertices is specified by ${N \choose 2}$ variables $x_{ij}$
with $x_{ij}=1$ if there is an edge between vertices $i$ and $j$. 
The access to $x_{ij}$ is by queries to a black box.
The task is to find if the graph $G$ contains a triangle.

\begin{theorem}
\cite{MSS}
{\bf Triangle finding} can be solved with $O(N^{1.3})$ quantum queries. 
\end{theorem}

The construction \cite{MSS} uses element distinctness as a subroutine 
in a clever two-level construction reminiscent of the $O(N^{3/4})$ 
algorithm for element distinctness in section \ref{sec:ed}.
Another problem for which element distinctness is useful as a subroutine is

{\bf Matrix product.}
Three $N\times N$ Boolean matrices $A, B$ and $C$ are specified by variables
$a_{ij}$, $b_{ij}$, $c_{ij}$, $n^2$ variables per matrix. 
The access to the variables is by queries to a black box.
The task is to find if $AB=C$, with the arithmetic operations modulo 2.

\begin{theorem}
\cite{BS}
{\bf Matrix product} can be solved with $O(N^{5/3})$ quantum queries.
\end{theorem}

\section{Conclusion and open problems}

In this column, we reviewed some of quantum 
algorithms for search problems: Grover's search,
amplitude amplification, their applications to NP-complete problems,
element distinctness and finding local and global minima,
and improved quantum search algorithms using quantum walks.

There are other interesting results that share similar ideas
or use the number of queries as the complexity measure.
To mention a few, \cite{BHT} have constructed a quantum
algorithm for collision problem,
\cite{Counting,Grover98,NW} have given quantum algorithms for 
approximate counting, finding mean and median,
\cite{HNS} studied quantum complexity of searching
among $N$ ordered items and sorting and there is
a large amount of work on quantum lower bounds 
(e.g. \cite{BBBV,Beals,Ambainis00}).

We conclude with some related open problems.
\begin{enumerate}
\item
{\bf Complexity of graph problems.}
Complexity of several graph problems remains open in the query model.
First, can the $O(N^{1.3})$ query triangle algorithm be improved? The best lower 
bound for this problem is $\Omega(N)$ (folklore). Second, what is the 
query complexity of finding a matching in a bipartite graph $G$ with $N$ vertices
on each side, specified by $N^2$ variables? There is an $\Omega(N^{1.5})$ lower
bound but no quantum algorithm that uses $o(N^2)$ queries.
\item
{\bf Generalizing quantum walk algorithms.}
As we saw in section \ref{sec:apps}, amplitude amplification provides
an easy way to apply Grover's technique to various problem without 
going into details of Grover's search algorithm. 
Quantum walk results (theorems \ref{th:akr} and
\ref{th:ed}) share common proof ideas. Can we find
an generalization for these two results which would be as easy-to-use
as amplitude amplification?

Results of Szegedy \cite{Szegedy} (e.g. Theorem \ref{th:szegedy})
are a major advance in this direction.
\item
{\bf Space usage in element distinctness.}
Both known algorithms for element distinctness use considerable amounts of memory
which has caused some criticism \cite{RG}.
The $O(N^{3/4})$ algorithm of \cite{Buhrman01} stores values of $O(\sqrt{N})$ 
variables and the $O(N^{2/3})$ algorithm of \cite{Ambainis04} stores
$O(N^{2/3})$ variables. Is it possible to design a quantum algorithm that uses
less space? Or can we prove a time-space lower bound saying that there is no
algorithm with better space usage for the given number of queries? 
\item
{\bf Quantum-classical relations.}
The quantum speedups described in this column are polynomial rather than
exponential (as in Shor's factoring algorithm).
This is inherent for a wide class of problems.
Consider computing a total Boolean function $f(x_1,\ldots, x_N)$,
with the variables $x_1, \ldots, x_N$ given by a black box that answers queries
(as in most problems described in this column).
Let $D(f)$ be the number of queries needed to compute $f$ 
deterministically and $Q(f)$ be the number of queries needed
to compute $f$ by a (bounded-error) quantum algorithm. 
Then, $D(f)$ and $Q(f)$ are polynomially related: 
$D(f)=O(Q^6(f))$ \cite{Beals}. 

The open question is: what is the biggest possible 
gap between $D(f)$ and $Q(f)$?
The best known result is for Grover's search problem: 
$D(f)=N$, $Q(f)=\Theta(\sqrt{N})$, $D(f)=\Theta(Q^2(f))$.
Can we find $f$ with a bigger gap or improve the
$D(f)=O(Q^6(f))$ relation?

A similar problem is open if we consider $Q_E(f)$, the number of
queries needed by the best {\em exact} quantum
algorithm (an exact algorithm is one which gives correct answer 
with certainty, instead of probability $1-\epsilon$). 
Then, we know that $D(f)=O(Q_E^3(f))$ \cite{Midrijanis}
but there is no known example for which $Q_E(f)=o(D(f))$.
For more information on this topic, we refer the reader to
an excellent survey by Buhrman and de Wolf \cite{BW}. 
\end{enumerate}

\end{document}